\documentstyle[psfig,aps]{revtex}

\tightenlines

\def\beq{\begin{equation}}
\def\eeq{\end{equation}}
\def\bea{\begin{eqnarray}}
\def\eea{\end{eqnarray}}

\def\d4k{\frac{d^4K}{(2\pi )^4}}

\begin{document}
\draft

\title{Strange Quark Matter in Neutron Stars? -- New Results from Chandra 
and XMM}

\author{Markus H. Thoma, Joachim Tr\"umper, Vadim Burwitz}
\address{Max-Planck-Institut f\"ur extraterrestrische Physik, 
Giessenbachstra{\ss}e, 85748 Garching, Germany}

\date{\today}

\maketitle

\begin{abstract}

It has been predicted that quark and hybrid stars, containing strange
quark matter in their core, have a significantly smaller
radius than ordinary neutron stars. Preliminary X-ray observations of
isolated neutron stars indicated a surprisingly small radius consistent
with quark matter in the interior of the star. However, a new analysis
of the data led to a radius corresponding to an ordinary neutron star.
In the present talk we will discuss theoretical calculations of
the mass-radius relation for quark and hybrid stars, taking into account
medium effects in quark matter, and report on recent X-ray observations by
Chandra and XMM and their latest interpretation.

\end{abstract} 

\section{Introduction}

Neutron stars provide an unique possibility to study matter at ``low'' 
temperature and ultra-high density. The density in the interior of a
neutron star could exceed a few times nuclear density, maybe up to about 
ten times at the center. Hence one might expect new forms of matter, such
as hyperonic matter, pion or kaon condensates, or in particular
deconfined quark matter.

The measurement of the mass $M$ and the radius $R$ of neutron stars 
puts stringent constraints on the 
equation of state (EOS) of the matter inside of
the star. Since most, if not all, neutron stars have a mass of about $M=1.4 
\> M_\odot$\footnote{At least, all binary neutron stars, for which the mass
can be determined accurately, have masses close to $1.4\> M_\odot$. There are
indications for larger masses from X-ray observations, which, however,
suffer from a large uncertainty or are model dependent \cite{ref0}.}, 
it is especially important to determine the radius accurately.
The observed radius $R_\infty$ is related to the actual radius $R$ by
\beq
R_\infty = R/\sqrt{1-2GM/Rc^2}.
\label{eq1}
\eeq
A radius $R<10$ km corresponding to $R_\infty< 13$ km ($M=1.4 
\> M_\odot$) cannot be explained 
with a conventional hadronic EOS. Hence it would be a strong indication 
for a phase transition to a new form of matter in the interior, since 
a phase transition leads to a softening of the EOS,
allowing therefore for smaller radii.

Recently, 
Chandra and XMM have measured X-ray spectra of isolated neutron stars
which had been discovered with ROSAT.
These spectra give the surface temperature $T$ related to the observed
temperature by $T_\infty=T\sqrt{1-2GM/Rc^2}$.
Assuming black-body radiation and measuring the distance $d$,
the radius follows from
\beq
R_\infty = d\sqrt{f_\infty/\sigma 
T_\infty^4},
\label{eq2}
\eeq
where $f_\infty$ is the integrated flux (observed at a large distance from the
star) corresponding to the temperature of the black-body.  

In the case of the isolated neutron star RXJ1856, $R<6$ km has been obtained 
in this way \cite{ref1}, suggesting that this isolated neutron star is 
either a 
strange quark star or a hybrid star, as we will discuss below.
Other signatures for quark matter in neutron stars are e.g. the cooling rate
\cite{ref3} or the rotation frequency \cite{ref4}. In this talk we will 
concentrate on the implications of radius measurements using X-ray
spectra. We will show that the results from Chandra and XMM combined
with optical data actually do not require RXJ1856 to be a quark matter star.

\section{Quark Matter in Neutron Stars (Theory)}

The idea of the presence of quark matter in compact stars was 
proposed a long time ago \cite{ref5}. It turned out that a
Fermi gas of up-, down-, and strange quarks is the energetically
most favorable state of quark matter at high densities. Such a system could
be realized in two possible ways:

1. Strange quark matter could be absolutely stable, i.e. more strongly
bound than the ground state of normal nuclear matter, i.e. iron \cite{ref6}.
Then strange quark stars could exist as self-bound systems
without any hadronic matter around it, although a thin hadronic crust is 
conceivable \cite{ref7}.

2. Quark matter is not absolutely stable but exists at high densities
in the interior of a neutron star, where at a certain density the
transition from hadronic to quark matter takes place. Such an object
containing a quark matter core surrounded by a hadronic mantle is called a 
hybrid star \cite{ref8}.

Now we want to ask: What do quark star models predict for the radius of
the star? Using the EOS of the model within the Tolman-Oppenheimer-Volkoff
equation gives a characteristic mass-radius relation of the star model.
Different quark matter EOS have been studied. First only the MIT bag model
without \cite{ref9} and with perturbative corrections \cite{ref10}
have been applied. Recently, a number of other models, such as
the Effective Mass Bag Model \cite{ref11}, the NJL Model \cite{ref12},
the non-perturbative Dyson-Schwinger equation \cite{ref13}, 
a quasiparticle model using input from lattice QCD \cite{ref14},
the Hard-Dense-Loop approach \cite{ref15}, and color superconductivity 
\cite{ref16}, have been adopted.

As an example, we want to discuss the Effective Mass Bag Model, in which
medium effects in quark matter at zero temperature $T=0$ but finite
baryon density, i.e. finite quark chemical potential $\mu\neq 0$,
have been taken into account. One of the most important medium effects
considered in many fields of physics are effective masses generated
by the interaction of the particles, e.g. within a mean-field approach
or a quasiparticle Fermi gas.
Here we consider effective quark masses following from the interaction 
of the quarks in quark matter by gluon exchange. The effective quark mass 
is defined here as the zero momentum limit of the quark dispersion relation.
To lowest order perturbation theory, which should hold at ultra-high 
densities according to asymptotic freedom, the dispersion relation follows
from the one-loop quark self-energy in the Hard-Dense-Loop limit
\cite{ref11} as
\beq
m_q^*=\frac{m_q}{2}+\left 
(\frac{m_q^2}{4}+\frac{g^2\mu^2}{6\pi^2}\right)^{1/2},
\label{eq3}
\eeq
where the bare (current)
up- and down-quark masses can be neglected, i.e. $m_u=m_d\simeq 0$,
compared to typical values of the quark chemical potential
$\mu =300$ - 400 MeV. The bare strange quark mass, on the other hand,
which is of the order of $m_s\simeq 150$ MeV, has to be taken into account.
In the Effective Mass Bag Model the strong coupling constant $g$ is considered
as a free quantity, which parametrizes the medium effect. We consider
coupling constants from $g=0$, corresponding to no medium effect, up to
$g=4$, giving the maximum medium effect in our model. Assuming that the most
important effects of the interactions are included in the effective masses
and the bag constant, we use these effective masses within the bag model.

\begin{figure}
\centerline{\psfig{figure=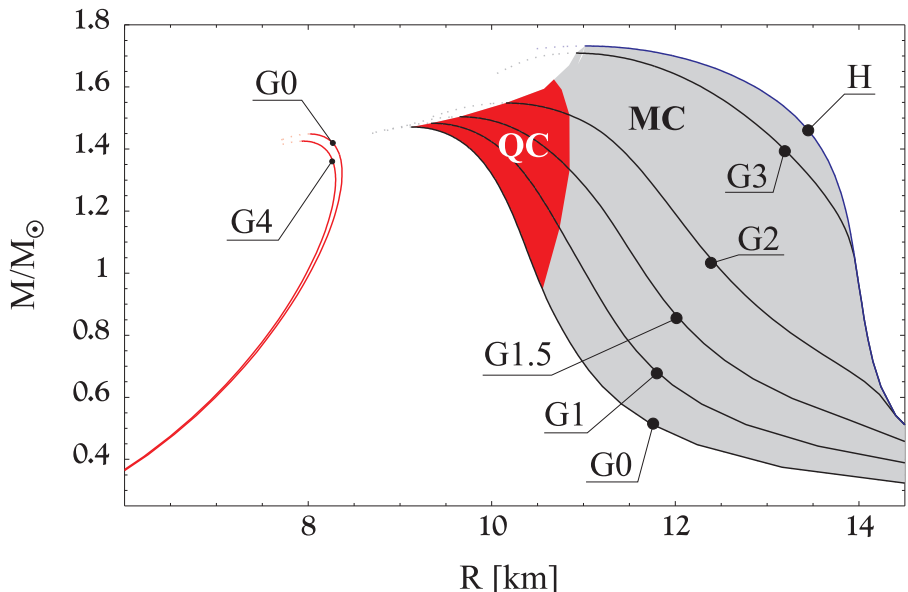,width=12cm}}
\caption{Mass radius relation for pure strange quark matter stars (left) 
and hybrid stars 
(right); G0: $g=0$ (no medium effect), ..., G4: $g=4$ (maximum medium 
effect); H: pure hadronic star, QC: star has a quark core, MC: star has a 
mixed core.}
\end{figure}

In this model the density per flavor is given by \cite{ref11}
\beq
\rho_q (\mu)=\frac{1}{\pi^2}\> [\mu^2-{m_q^*}^2(\mu)]^{3/2}.
\label{eq4}
\eeq
From the thermodynamic relations one finds the
pressure and energy density
\beq
\rho_q(\mu)=\frac{dp_q(\mu)}{d\mu}, \; \; \; \; \; \;  
\epsilon_q(\mu)+p_q(\mu)=\mu \, \rho_q(\mu)
\label{eq4a}
\eeq
The energy density and pressure of strange quark matter reads
\bea
\epsilon_{SQM}&=&\epsilon_u+\epsilon_d+\epsilon_s+\epsilon_e+B,\nonumber \\
p_{SQM}&=&p_u+p_d+p_s+p_e-B,\nonumber
\eea
where also the electron energy density ($\epsilon_e$) and pressure ($p_e$)
have been considered.

We have restricted ourselves to bag constants between $B^{1/4}= 165$ and
200 MeV. For lower values the phase transition to quark matter takes
place already at sub-nuclear density. In particular, absolutely stable 
strange quark matter is excluded in our model as it requires
$B^{1/4}<155$ MeV. For $B^{1/4}>200$ MeV we find no quark matter in our
star models. It should be noted that the effect of the effective quark
mass cannot be reproduced by altering $B$, $m_s$ or 
introducing $\alpha_s$-corrections.

Next we have solved the Tolman-Oppenheimer-Volkoff (TOV) equation numerically,
where the EOS enters only through the pressure as a function
of the energy density, $p=p(\epsilon)$. In this way we obtained the
mass-radius relation shown in Fig.1. Note the different qualitative behavior
of strange quark stars and hybrid stars in this plot, where the latter
behave similar as hadronic stars.

In the case of strange quark stars, which are located at radii $R<9$ km, 
we observe that medium effects are negligible \cite{ref11}.
In the case of hybrid stars, on the other hand, medium effects play
a crucial role. Small medium effects, corresponding to $g<2$, 
imply a quark core in the center of the star whereas larger values
allow at most a mixed phase in the center. This is caused by the fact 
that a large effective quark mass enhances the energy density of quark matter
making its presence in the star less likely, as shown in Fig.2.

\begin{figure}
\centerline{\psfig{figure=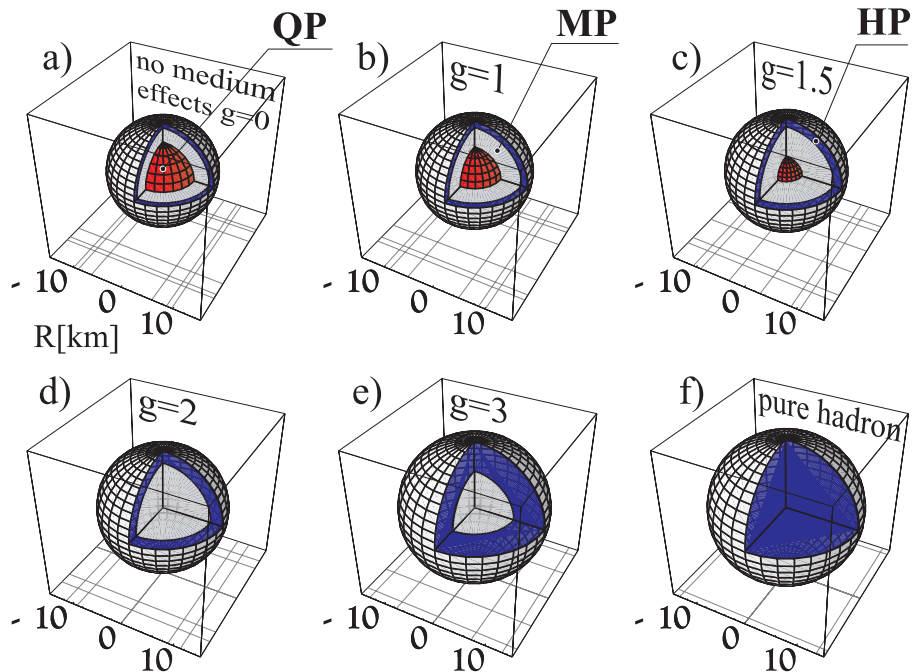,width=12cm}}
\caption{Schematic gross structure of a 1.4 $M_\odot$ hybrid star
(QP: quark phase, MP: mixed phase, HP: hadronic phase).}
\end{figure}

We have used the quark matter EOS for the quark core and for 
the quark component of the mixed phase, which builds a substantial part 
of the star, and
a hadronic EOS for the hadronic component of the mixed phase and for the
hadronic mantle. In order to test the sensitivity of our results to the
hadronic EOS, we employed four different types of relativistic mean-field 
models including hyperons. It turns out that, independently of the hadronic
model used, a neutron star with a quark matter core is typically 
20 - 30\% smaller than a pure hadronic star. A pure hadronic star
has a radius larger than 13 km (for $M=1.4\> M_\odot$), while
hybrid stars with a quark matter core have radii between 9 and 11
km.

The mass-radius relation
of strange quark stars, using the MIT bag model without effective 
quark masses, and of hadronic stars, using different hadronic EOS, was
also considered 
in Ref.\cite{ref17}. It was also found that the strange quark stars are
typically smaller than hadronic stars, although some overlap between
the radii of those stars is possible depending on the mass of the star
and the choice of the strange quark matter parameters, in particular if
a soft hadronic EOS is used. However, in recent investigations
of perturbative \cite{ref18} and Hard-Dense-Loop corrections \cite{ref15}
to the Bag Model, it has been shown that
the presence of ultraviolet divergences leads to a strong
dependence of the mass-radius relations of strange quark  
and hybrid stars on the renormalization scale, which had to be introduced
to remove the divergences. For example, in Ref.\cite{ref18} the choice
of the renormalization constant of $2\mu$ gives a maximum radius of
about 6 km with a maximum mass of about 1 $M_\odot$, while the choice $3\mu$ 
yields a maximum radius of more than 12 km and a maximum mass of about
2 $M_\odot$. 

Using the NJL model for the quark matter EOS, neither a strange quark star
nor a hybrid star with a quark matter core exists. Only a mixed phase 
in center is possible for certain parameter choices \cite{ref12}, similar
as in the Effective Mass Bag Model in the case of large values of the
strong coupling constant. 

Finally, recent studies of neutron stars, based on a hadronic model 
with strongly interacting (attractive) hyperons \cite{ref19} and with 
a kaon condensate \cite{ref20} showed a radius possibly as small as about
8 km for a 1.4 $M_\odot$ star which is comparable to hybrid and
even to strange quark stars.

From the various results discussed above, we conclude that
there is a large theoretical uncertainty for extracting useful constraints
on the EOS from the mass-radius relation. Hence it would be desirable
to have a reliable EOS for quark as well as hadronic matter
within a single approach. In principle such a possibility could be provided by
lattice QCD simulations, which, however, so far are not possible at
finite density and zero temperature, although some progress has been 
made recently \cite{ref21}.
 
\begin{figure}
\centerline{\psfig{figure=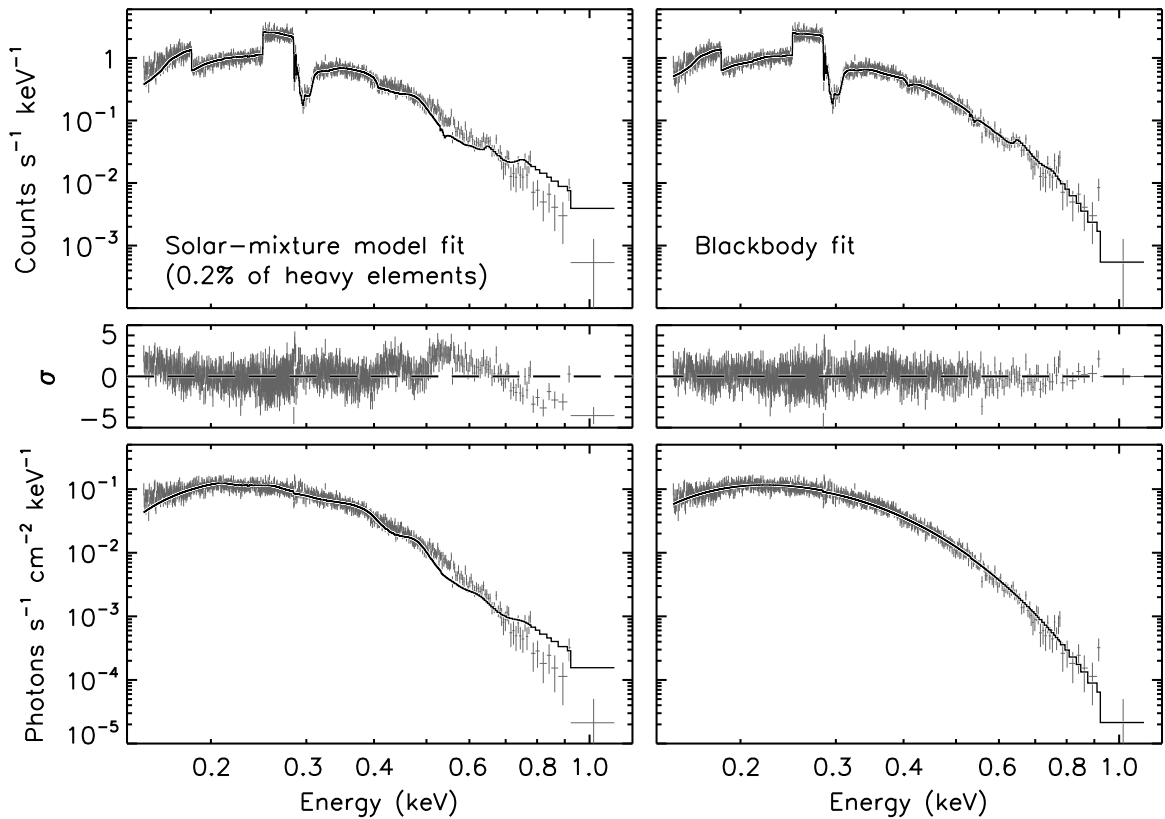,width=16cm}}
\caption{Counts and spectra of RXJ1856 measured by Chandra and compared to a 
solar-mixture model (left) and a black-body (right). Figure taken from
Ref.[22].}
\end{figure}

\section{Chandra and XMM Observations}

Neutron stars are X-rays sources. The X-rays are emitted
either from the hot surface ($T\sim 10^6$ K) or are due to 
a nonthermal radiation from the co-rotating magnetosphere
in the case of a pulsar. 
These radiation modes are distinguishable by their spectra, namely
a black-body or a power law spectrum, respectively. So far seven
isolated neutron stars with a thermal X-ray spectrum have been
discovered with ROSAT. RXJ1856 is the brightest of these objects. It shows  
no indications of pulsation \cite{ref22}. It is important that
RXJ1856 has also been detected by the Hubble space telescope 
at a magnitude of $V\simeq 26$ mag \cite{ref23}. From optical measurements
of the parallax the distance has been determined as
$d=117\pm 12$ pc \cite{ref24}.

Chandra and XMM have measured high-resolution X-ray spectra of RX1856.
These spectra agree perfectly with black-body spectra and show no 
indication of lines which are predicted by existing photospheric 
models \cite{ref27}. In Fig.3 the measured counts and the spectra,
corrected for the acceptance of the detectors, are shown.  
In the left panels these quantities are compared to a solar-mixture model 
containing 0.2\% of heavy elements given by the solid line. The right
panels, on the other hand, show the comparison with a black-body spectrum.
Whereas the latter agrees perfectly with the data, the solar-mixture model is 
clearly ruled out. Also a broad-band spectral feature, e.g. caused by a fast 
rotation are unlikely. However, a strong magnetic field may lead to a 
splitting and smearing of the lines by field variations across the surface,
which could result in a featureless spectrum similar to a black-body.

Assuming black-body radiation in the X-ray regime,
a temperature of $kT_\infty^X=63\pm 0.5$ eV is inferred 
from these measurements.
Together with the known distance an amazingly small 
radius of
\beq
R_\infty^X = 4.3\pm 0.2 {\rm km}\; (d/120 {\rm pc})
\label{eq5}
\eeq
is found. From this radius it was concluded
that RXJ1856 has to be a strange quark star without a hadronic crust
explaining also the absence of lines \cite{ref1,ref26}.

\begin{figure}
\centerline{\psfig{figure=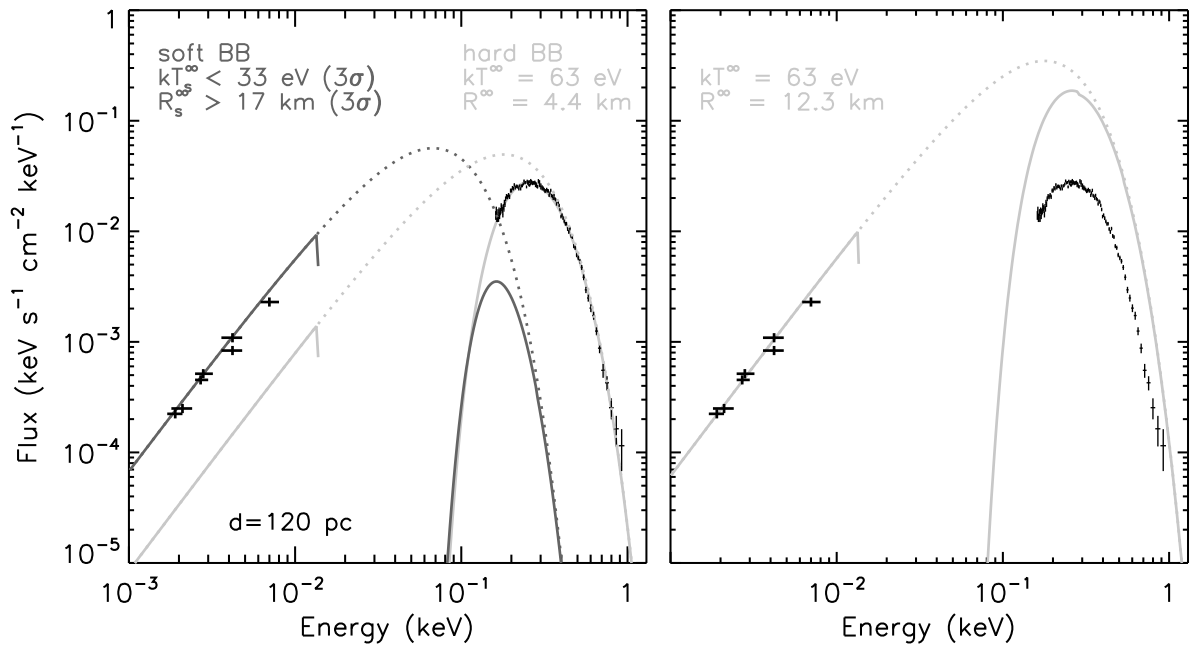,width=16cm}}
\caption{Possible explanations of X-ray and optical fluxes, 
assuming a two-temperature spectrum (left) and a uniform temperature
distribution together with a grey-body X-ray spectrum (right).
The black crosses are the data, the solid and dotted lines the 
extrapolated spectra for the two different models. Figure taken from
Ref.[22].}
\end{figure}

However, this interpretation does not explain the 
optical data (see Fig.4). 
Extrapolating the X-ray spectrum to the optical regime,
one observes that the optical flux exceeds the extrapolated one by 
a factor of about 7. A possible explanation would be that there are
two different sources of the optical and the X-ray emission: the X-ray
spectrum is produced by a hot spot on the surface, while the optical data 
come from the entire surface. Then using the constraint that the soft 
component should not distort the X-ray spectrum of the hard component, 
a surface temperature of
$T_\infty^{opt}<33.6$ eV follows, which results in a radius
of $R_\infty^{opt}>16.3$ km ($d$/120 pc) \cite{ref22,ref27}. 
This solution is shown in  
the left panel of Fig.4, where the optical black-body spectrum is 
constructed in way that the extrapolation to the X-ray regime falls below
the measured data. The difference between the X-ray data and the extrapolation
of these data to the optical regime (dotted grey curve) arises by taking
interstellar absorption into account. 

However, the two-temperature 
explanation is questionable since an upper limit of 1.5\% 
for periodic
variations in the X-ray data has been found by XMM. This puts a severe 
constraint on the relative alignment of the rotational axis and the line of 
sight.
Hence, a more reasonable explanation 
may be based on the assumption of a uniform surface temperature distribution
together with a suppressed X-ray emission in the X-ray band \cite{ref22}. 
As discussed above, a suppression of the X-ray flux
by a factor 0.15 is needed for explaining the optical data.
However, the shape of the observed black-body spectrum should not
be changed. Such a spectrum corresponds to a ``grey-body'' 
spectrum. Fitting the optical spectrum, assumed to be a black-body spectrum,
to the X-ray spectrum taking into account the suppression factor is shown
in the right panel of Fig.4. 

A suppression of the black-body radiation is expected if there were a
condensed matter surface consisting of iron or hydrogen \cite{ref30,ref31}. 
Such a surface is expected to
form at low temperatures and strong magnetic fields. 
For RXJ1856
this could occur at magnetic fields above about $10^{12}$ G \cite{ref32}. 
Unfortunately,
the magnetic field of RXJ1856 is not known. However, even stronger magnetic
fields of neutron stars have been observed. 

The electron plasma frequency in a condensed matter surface is estimated to 
be $\hbar \omega_{pl} \simeq 0.66\> 
B_{12}^{3/5}$ keV ($B=B_{12}\times 10^{12}$ G),
which is much larger than the typical photon energy
$kT\sim E_\gamma$. Hence, photons cannot be excited easily within the
metallic surface \cite{ref33}, resulting in a high reflectivity of the surface.
Then the spectrum can be described by a black-body spectrum
times an energy dependent absorption factor,
$J(E_\gamma)=\alpha_X(E_\gamma)\> J_{BB}(E_\gamma)$. The best fit, 
corresponding to
a surface temperature $kT_\infty = 54$ eV, yields an even better fit than
a pure black-body. This leads to an observed radius
of $R_\infty \geq 13.3$ km \cite{ref22}. 
(The equal sign holds if there is no suppression
of the emission in the optical regime.). For a $1.4 M_\odot$ star an actual 
radius of
$R>10.3$ km follows which is consistent with the assumption of a normal 
neutron star. Similar models and conclusions have been discussed
by Walter in \cite{ref34} and by Zane, Turolla and Drake in \cite{ref35}.

We conclude that there is no need for exotic
matter in the interior of neutron stars for explaining the observations so 
far. Either a two-temperature model, a strong magnetic field, or 
a grey-body X-ray emission are capable of explaining the observed
data in accordance with a conventional hadronic EOS. In addition, the various
theoretical predictions of the mass-radius relation 
suffer from a large uncertainty, which does not allow one to distinguish
between quark matter and hadronic matter if the radius is not clearly below
8 km. Hence a better understanding of the EOS would be desirable.
Also, it appears to be equally important to understand the surface 
properties of neutron stars better.

After presenting this talk, new evidence appeared that smearing of the
spectrum by a strong magnetic field is the most likely explanation.  
First, a strong overall suppression of the blackbody spectrum by a
factor of 7 appears to be difficult to explain. New investigations
show only a suppression by a factor 2 to 3 \cite{ref35}. Second,
an iron condensate surface, as discussed by Lai \cite{ref32}, is not
supported by other investigations \cite{ref36}. At least, models
describing an iron condensate surface are still quite crude
\cite{ref35}. A hydrogen condensate, 
however, requires a probably unrealistically high magnetic field 
($B>5\times 10^{13}$ G).
A strong magnetic field (up to $10^{13}$ G), on the other hand, 
as it might exist at RXJ1856
leads to a splitting of the atomic levels in the X-ray regime into
numerous Landau levels. A field variation between the pole and the equator
of the star by a factor of two, as expected in the case of a dipole field,
results in a smearing of the dense lines such that the lines cannot be 
resolved anymore. In this way a X-ray spectrum arises in accordance with the
one measured by Chandra and XMM. A strong magnetic field has also
been discussed recently as the most likely interpretation of the observed
spectra of the isolated neutron stars RXJ1308 \cite{ref37} and RXJ0720 
\cite{ref38}. Combining such a X-ray spectrum with the optical
spectrum, one finds a black-body radius $R_\infty >17$ km, indicating a stiff
equation of state, which would exclude a strange quark matter star or
even hybrid star. This is a rather conservative lower limit for the radius
since a black-body emitter is the most efficient radiator.

\end{document}